\documentclass[12pt,a4paper]{article}

\usepackage{multirow}
\usepackage{t1enc}
\usepackage[utf8x]{inputenc}
\usepackage{amsmath, amssymb, amsfonts, multirow}
\usepackage{graphicx}
\usepackage{grffile}
\usepackage{float} 
\usepackage{fullpage}
\usepackage{verbatim}
\usepackage{mathtools}
\usepackage{fancyhdr}

\usepackage{hyperref}
\usepackage{color}

\usepackage{url}

\usepackage{caption}

\usepackage{adjustbox}

\usepackage{array}

\usepackage{authblk}

\title{Phase space volume scaling of generalized entropies and anomalous diffusion scaling governed by corresponding non-linear Fokker-Planck equations}

\begin{document}



\author[1]{D\'{a}niel Cz\'{e}gel}
\author[2]{S\'{a}muel G Balogh}
\author[3]{P\'{e}ter Pollner}
\author[3]{Gergely Palla}
\affil[1]{Evolutionary Systems Research Group, MTA Centre for Ecological Research, Hungarian Academy of Sciences, H-8237 Tihany, Hungary}
\affil[2]{Dept. of Biological Physics, E\"{o}tv\"{o}s University, H-1117 Budapest, Hungary}
\affil[3]{MTA-ELTE Statistical and Biological Physics Research Group, Hungarian Academy of Sciences, H-1117 Budapest, Hungary}

\maketitle

\vspace{2cm}

\section*{Abstract}

Many physical, biological or social systems are governed by history-dependent dynamics or are composed of strongly interacting units, showing an extreme diversity of microscopic behaviour. Macroscopically, however, they can be efficiently modeled by generalizing concepts of the theory of Markovian, ergodic and weakly interacting stochastic processes. In this paper, we model stochastic processes by a family of generalized Fokker-Planck equations whose stationary solutions are equivalent to the maximum entropy distributions according to generalized entropies. We show that at asymptotically large times and volumes, the scaling exponent of the anomalous diffusion process described by the generalized Fokker-Planck equation and the phase space volume scaling exponent of the generalized entropy bijectively determine each other via a simple algebraic relation. This implies that these basic measures characterizing the transient and the stationary behaviour of the processes provide the same information regarding the asymptotic regime, and consequently, the classification of the processes given by these two exponents coincide.

\section*{Introduction}

Real world processes are often characterized by the presence of a large number of interacting phenomena at multiple time- or length scales \cite{bar2013beyond}, and thus, they are usually described by stochastic models that are strongly interacting or history-dependent \cite{shibata1977generalized, frank2003note, hanel2013generalized, corominas2015understanding}. A way to understand and classify these processes in terms of stationary and non-stationary probability densities is to generalize the concepts of statistical mechanics that already proved to be very powerful for describing weakly interacting, ergodic and Markovian systems \cite{tsallis2009introduction, beck2009generalised}. One such concept is entropy, which assigns a likelihood to macrostates, that is, to stationary distributions over microstates. Maximizing this likelihood, possibly in the presence of external constraints, yields the most probable stationary distribution characterizing the system, called the Maximum Entropy (MaxEnt) distribution, which plays a key role in describing the stationary behaviour of stochastic systems. For example, the Boltzmann-Gibbs entropy form, $S_{BG}=-\sum_i p_i \ln p_i$, where $i$ runs over the microstates, follows from the assumption that the system realizations are independent and distinguishable \cite{pathria1972statistical}.

In general, however, the realizations are not independent. Instead, their interaction can be macroscopically modelled by a corresponding entropy functional, which in principle can take infinitely many different forms. Similarly to the theory of renormalization group describing critical phenomena, an apprehensive characterization of these entropies can be made by observing what are the relevant and irrelevant parameters as we approach infinite system size \cite{stanley1999scaling}. Axiomatic considerations suggest that the asymptotic scaling of the generalized entropy forms with phase space volume provides a meaningful classification of the entropies.
This classification is based on the fundamental result by Hanel and Thurner\cite{hanel2011comprehensive} about the entropy functionals 
$S[p]$ that can be written as a sum of a pointwise function over 
microstates
\begin{equation}
S[p]=\sum_i g(p_i).
\label{discrete_S}
\end{equation}
As they showed, the 
first three Shannon-Khinchin (SK1-SK3) axioms \cite{shannon1948mathematical,khinchin1957mathematical}
\begin{itemize}
\item[SK1] $S$ is continuous in $p$,
\item[SK2] $S$ is maximal for the uniform distribution, $p_i\equiv 1/W$,
\item[SK3] $S$ is invariant under adding a zero-probability state to the system, $S(p_1,\dots, p_W)=S(p_1,\dots, p_W,p_{W+1}=0)$,
\end{itemize}
permit only the following asymptotic scaling relation for any entropic forms:
\begin{equation}
\lim_{W\to \infty}\frac{S[p_i\equiv (\lambda W)^{-1}]}{S[p_i\equiv (W)^{-1}]}=\lambda^{1-c},
\label{lambda_scaling}
\end{equation}
or, equivalently, in terms of $g$,
\begin{equation}
\lim_{p\to 0} \frac{g(zp)}{g(p)}=z^c, 
\label{g_scaling}
\end{equation}
with $0<z<1$ and $0<c\leq 1$.

Hence, the scaling exponent $c$ can be used to parametrize the equivalence classes of the generalized entropy forms. For example, the Boltzmann-Gibbs entropy, where $g(p)=-p\ln p$, is corresponding to $c=1$, whereas the Tsallis entropy \cite{tsallis2009introduction} $S[p]=\frac{1-\sum_i p_i^q}{q-1}$, 
with $0<q\leq1$ is corresponding to $c=q$. 
Consequently, each such equivalence class can be represented by a Tsallis entropy. Note that the fourth Shannon-Khinchin axiom $S(p_{AB})=S(p_A)+\langle S(p_{B|A}) \rangle_A$ is not considered in this analysis, therefore, the entropy of a joint distribution $p_{AB}$ is not always decomposable to the entropy of the marginal $p_A$ and the entropy of the conditional distribution $p_{B|A}$, averaged over $p_A$.

These considerations suggest that the asymptotic exponent $c$ provides a measure of deviation from ergodic, uncorrelated and Markovian systems regarding its stationary behaviour. Our main motivation here is to understand how this exponent relates to similar macroscopic measures which are, however, characterizing the non-stationary behaviour of the system. One of the main approaches to model the non-stationary behaviour of stochastic processes macroscopically is through partial differential equations governing the time evolution of the probability density $p(x,t)$, called Fokker-Planck equations (FPE)\cite{van2007stochastic, toral2014stochastic,frank2005nonlinear}. Once specified, the underlying microscopic rules completely determine the form of the FPE. For example, the assumption of memoryless, Gaussian noise and short range interaction between the units gives rise to linear FPEs in the form of 
\begin{equation}
\partial_t p(x,t)=-\partial_x(p(x,t)f(x,t))+\partial_x(D(x,t)\partial_x p(x,t)),
\label{linFPE}
\end{equation}
where $f(x,t)$ and $D(x,t)$ are called the drift and diffusion coefficients, respectively. If the diffusion coefficient is constant ($D(x,t)\equiv D$) and the drift is proportional to the spatial derivative of some time-independent external potential $u(x)$, i.e., $f(x,t)=-D\beta\partial_x u(x)$, (\ref{linFPE}) simplifies to
\begin{equation}
\partial_t p(x,t)=D\beta\partial_x (p(x,t)\partial_x u(x))+ D \partial_x^2 p(x,t).
\label{simple_linFPE}
\end{equation}
Specifically, in the presence of no external potential, (\ref{simple_linFPE}) becomes
\begin{equation}
\partial_t p(x,t)=D \partial_x^2 p(x,t).
\label{simple_diffusion}
\end{equation}

Such non-stationary processes can be classified phenomenologically by the scaling of the spread of $p(x,t)$ over time. This can be phrased mathematically as the invariance of $p(x,t)$ under appropriate rescaling of space and time\cite{bouchaud1990anomalous, dubkov2008levy}:
\begin{equation}
p(x,t)=\tau^{-\gamma}p\left(\frac{x}{\tau^{\gamma}},\frac{t}{\tau}\right),
\label{anom_diff_0}
\end{equation}
where the scaling factor $\tau^{-\gamma}$ of the space coordinate keeps the probability density invariant when the timescale is changed as $t\to \frac{t}{\tau}$. 
In general, (\ref{anom_diff_0}) is satisfied only in the asymptotic limit, i.e., when $p(x,t)\to 0$. Nevertheless, this scaling relation, parametrized by $\gamma$, classifies the governing dynamics described by (\ref{gFPdef}). For example, (\ref{simple_diffusion}) falls into the equivalence class $\gamma=1/2$.

Non-stationary stochastic processes that are characterized by $\gamma\neq \frac{1}{2}$ are termed as anomalous diffusion processes. There are two main types of microscopic rules that can lead to anomalous diffusion of the probability density. In one case, the trajectories of the individual units (e.g., particles) remain to be uncorrelated and Markovian, however, other stochastic properties of these trajectories deviate from those of standard Brownian motion\cite{metzler1999anomalous,metzler2000random,klages2008anomalous,dubkov2008levy}. Typically these deviations stem from the fact that either the waiting time distribution between successive jumps or the jump length distribution is characterized by having infinite variance or mean. The corresponding FPEs usually include fractional derivatives, hence, these processes are termed as fractional dynamics. We do not consider this type of processes in the rest of the paper. Instead, we focus exclusively on the other type of processes that can lead to anomalous diffusion: the case in which the dynamics of the units is correlated or non-Markovian\cite{stariolo1994langevin, borland1998microscopic,frank2001langevin, curado2003derivation, frank2003note, frank2005nonlinear,chavanis2008nonlinear, tsallis2009introduction, souza2017thermodynamic}. One way of modelling macroscopically such systems is through FPEs in which the diffusive term, $\partial_x^2 p$, is replaced by $\partial_x^2 F[p]$, where $F[p]$, called the effective density, is a given function of the probability density $p$, which is either derived from microscopic rules or simply defined based on other macroscopic arguments\cite{plastino1995non, borland1998microscopic, frank2005nonlinear,chavanis2008nonlinear}. According to the above, in the following we consider non-linear FPEs that generalize (\ref{simple_linFPE}) as
\begin{equation}
\partial_t p=D\{\beta \partial_x(p\partial_x u )+\partial_x^2 F[p]\}.
\label{gFPdef}
\end{equation}
Nonlinear FPEs were used in modelling a variety of phenomena in physical, biological and social sciences, such as diffusion in porous media \cite{bouchaud1990anomalous,spohn1993surface}, surface growth process \cite{spohn1993surface}, stellar dynamics \cite{chavanis2003generalized}, bacterial chemotaxis \cite{chavanis2008nonlinear} and financial transactions \cite{borland2002option}.

As we show later, for a given effective density $F[p]$, the asymptotic anomalous diffusion exponent $\gamma$ can be determined. Therefore, similarly to the phase space volume scaling exponent $c$ of the entropy, the anomalous diffusion exponent $\gamma$ might also indicate the deviation of the underlying system from being uncorrelated and Markovian. This specifies the goal of this paper, which is to investigate the relation between these two exponents, $c$ and $\gamma$, macroscopically characterizing the stationary and non-stationary regime of the process, respectively.
In order to relate entropies to FPEs, in this paper we consider continuous entropy forms, which, analogously to (\ref{discrete_S}), are assumed to be written as \cite{martinez1998nonlinear, frank1999nonlinear, chavanis2008nonlinear}
\begin{equation}
S[p(u)]=\int g\left(p(u)\right ) \mathrm{d}u,
\label{S_g_form}
\end{equation}
where $g$ is asymptotically characterized by (\ref{g_scaling}), $u=u(x)$ is a time-independent scalar function of the space coordinate $x$ (e.g., a potential), and the integration is performed over the range of $u(x)$. The definition given in (\ref{S_g_form}) provides a very general form, and special cases of this entropy functional have already been applied in studies of statistical mechanics of special relativity \cite{kaniadakis2002statistical}, chemotaxis of biological populations \cite{chavanis2008nonlinear}, stellar dynamics and two dimensional turbulence \cite{chavanis2003generalized}.

For the sake of consistency between the description of these two regimes, similarly to Refs.\cite{plastino1995non, tsallis1996anomalous, martinez1998nonlinear, frank1999nonlinear, schwammle2007general}, we consider cases where the stationary solution of the FPE, given by (\ref{gFPdef}), equals to the maximum entropy distribution according to the generalized entropy.
It is instructive to see how the consistency criterion specified above applies to the most well-known case, the Boltzmann-Gibbs entropy, $S=-\int \mathrm{d}u \ p(u) \ln p(u)$.
In this case, the MaxEnt distribution restricted by a constraint on the expected value of $u$ takes the form of $p(u)=\frac{1}{Z} e^{-\beta u}$. However, this is also equivalent to the stationary solution of the FPE describing ordinary diffusion in the presence of some external potential $u(x)$, given in (\ref{simple_linFPE}),
which is a special case of (\ref{gFPdef}) with $F[p]=p$. Setting zero net flux at the boundaries yields $p(u)=\frac{1}{Z} e^{-\beta u(x)}$.

\section*{Results}

\begin{figure}
  \centering
    \includegraphics[width=0.7\textwidth]{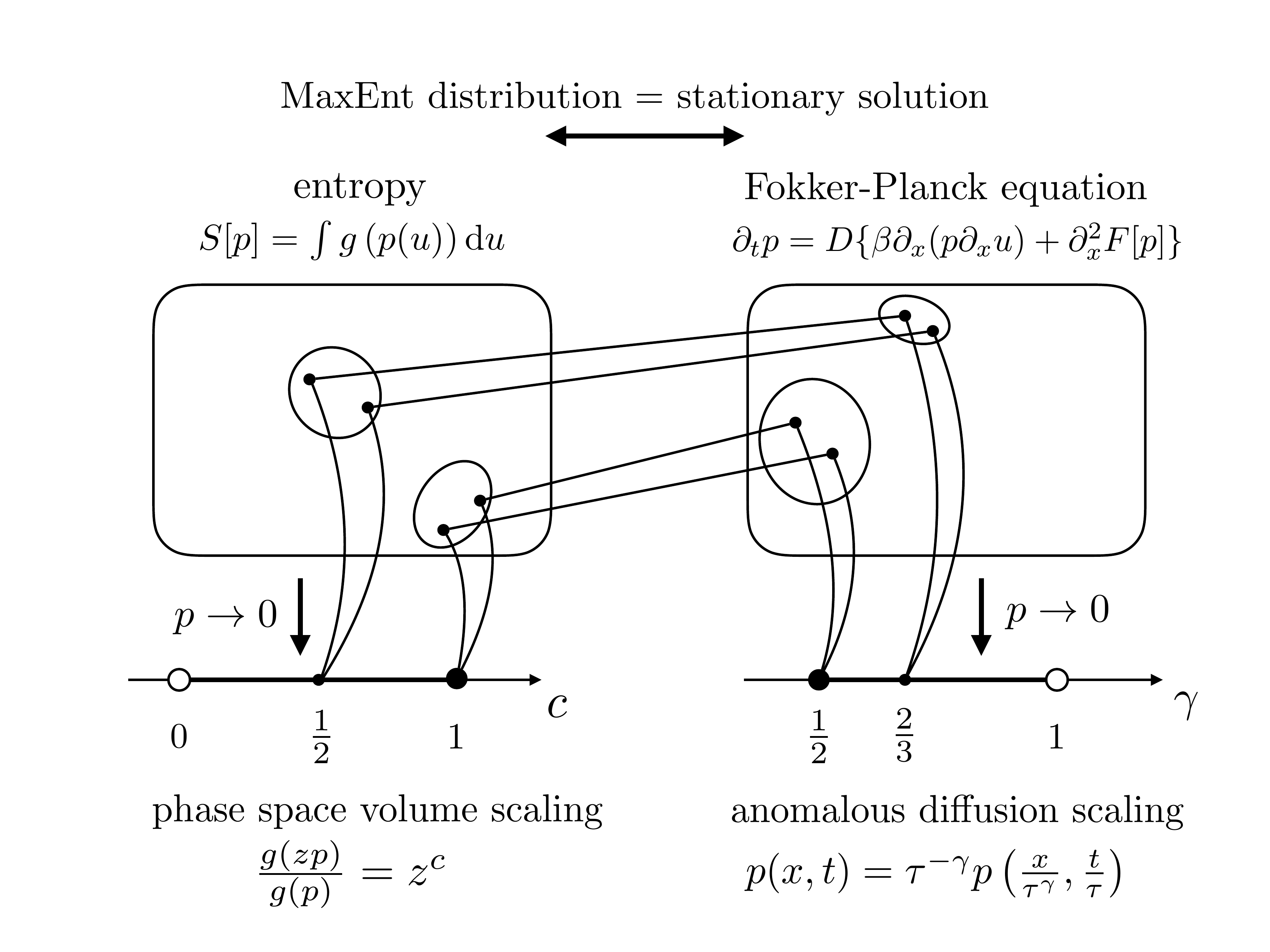}
      \caption{Summary of the results presented in this paper. We consider macroscopic descriptions of stochastic processes where the stationary and non-stationary regime are related by identifying the MaxEnt distribution with the stationary solution of the Fokker-Planck equation. We show that under these assumptions, two of the most frequently discussed asymptotic classifications of stochastic processes, one based on the scaling of the entropy with the phase space volume and the other based on the anomalous diffusion scaling described by the Fokker Planck equation, coincide. Thus, for any $c\in (0,1]$, the corresponding equivalence class of entropies (represented by an ellipse on the left) bijectively determines an equivalence class of Fokker-Planck equations (shown by an ellipse on the right), in which the anomalous diffusion scaling exponent $\gamma\in[\frac{1}{2},1)$ is constant.}
\label{fig:summary}
\end{figure}

Based on the above, the Boltzmann-Gibbs entropy, belonging to the entropy class $c=1$, is corresponding to the Fokker-Planck equation describing simple diffusion, which in turn is a member of the anomalous diffusion scaling class $\gamma=1/2$. A natural question arising based on this observation is the following: Does every entropy belonging to the $c=1$ universality class correspond to a generalized Fokker-Planck equation from the anomalous diffusion class $\gamma=1/2$? And does every generalized Fokker-Planck equation belonging to the class $\gamma=1/2$ correspond to an entropy belonging to the $c=1$ class? In other words, does $c=1$ and $\gamma=1/2$ give rise to the same equivalence class, therefore, bijectively determine each other? 
In this paper we show that this is true not only for $c=1$ and $\gamma=1/2$, but for every $c\in (0,1]$ and $\gamma\in[\frac{1}{2},1)$, where the exponents $c$ and $\gamma$ are connected by a simple algebraic relation. This implies that the asymptotic scaling of generalized entropies with phase space volume and the asymptotic anomalous diffusion scaling of the corresponding generalized Fokker-Planck equation classify the processes in the same way, and consequently, they provide the same information about their asymptotic behaviour. In Fig.\ref{fig:summary}. we show a schematic illustration of the above concept. 
Our result also provides an asymptotic generalization of the relation derived by Tsallis and Bukman \cite{tsallis1996anomalous} between $c$ and $\gamma$ for the class of Tsallis entropies.

In order to derive a relationship between the asymptotic exponents $c$ and $\gamma$, let us first consider the MaxEnt distribution corresponding to entropies given in the form of (\ref{S_g_form}). By following a variational principle approach and taking into account the normalization and expected value constraints we can write
\begin{equation}
\frac{\delta}{\delta p} \int \mathrm{d}u \left(g(p(u))-\lambda_0 p(u) -\lambda_1 up(u)\right)=0,
\label{general_deltaS}
\end{equation}
where the constants are omitted for simplicity and the $\lambda_0$ and $\lambda_1$ Lagrange multipliers are introduced for fixing the zeroth and first moment, respectively. From (\ref{general_deltaS}) we obtain
\begin{equation}
u=\frac{g'(p)-\lambda_0}{\lambda_1}=\Lambda(p),
\label{gen_log}
\end{equation}
where $\Lambda(p)$ is the inverse of the MaxEnt distribution corresponding to the entropy defined by $g(p)$. In case of the Boltzmann-Gibbs entropy the inverse of the MaxEnt distribution is given by the (appropriately shifted and rescaled) logarithm function, $\Lambda_{\rm BG}(p)=-\lambda_1^{-1}(\ln p+\lambda_0+1)=-\beta^{-1}(\ln p + \ln Z)$. Therefore, $\Lambda(p)$ is usually referred to as the generalized logarithm for any entropy in general \cite{tsallis2009introduction,hanel2012generalized}.

Based on a given entropy $S[g(p)]$ and the corresponding generalized logarithm $\Lambda(p)$, our next step is to find the related generalized Fokker-Planck equations in the form of (\ref{gFPdef}), 
where $F[p]$ is chosen such that the stationary solution of the equation becomes equivalent to the MaxEnt distribution of the entropy. 
By replacing $u$ with $\Lambda(p)$ in (\ref{gFPdef}) according to (\ref{gen_log}) we obtain that the stationarity condition $\partial_t p=0$ is fulfilled if
\begin{equation}
\partial_x^2F[p]=-\beta\partial_x(p\partial_x \Lambda(p)).
\label{F1}
\end{equation}
The expression $p\partial_x \Lambda(p)$ in the r.h.s. of (\ref{F1}) can be rewritten using the chain rule $\partial_x= (\partial_x p) \partial_p$ as
\begin{equation}
p\partial_x \Lambda(p)=(\partial_x p) p\partial_p \Lambda(p),
\end{equation}
which in turn is equivalent to
\begin{equation}
(\partial_x p) p\partial_p \Lambda(p)=(\partial_x p) \partial_p \left(\int_0^p q\partial_q \Lambda(q)\mathrm{d}q+C(x,t)\right)=\partial_x\left(\int_0^p q\partial_q \Lambda(q)\mathrm{d}q+C(x,t)\right),
\label{F2}
\end{equation}
where $C(x,t)$ is an arbitrary function which is independent of $p$. Substituting (\ref{F2}) into (\ref{F1}) yields
\begin{equation}
\partial_x^2F[p]=-\beta\partial_x^2\left(\int_0^p q\partial_q \Lambda(q)\mathrm{d}q+C(x,t)\right),
\end{equation}
which gives the general expression for the effective density $F$ as
\begin{equation}
F[p]=-\beta\int_0^p q \partial_q\Lambda(q) \mathrm{d}q+\tilde{C}=\frac{-\beta}{\lambda_1}\int_0^p q \partial_q^2 g(q) \mathrm{d}q+\frac{\tilde{C}}{\lambda_1},
\label{gen_F}
\end{equation}
where $\tilde{C}=-\beta^{-1}C(x,t)+a(t)+b(t)x$ is an arbitrary function that is constant in $p$. The obtained relation (\ref{gen_F}) between $F$ and $g$ has already been established in an implicit form in Refs.\cite{martinez1998nonlinear,chavanis2008nonlinear}.
In the following, we assume that $F$ has no explicit space- or time-dependence, consequently, it is defined by $g$ up to an additive constant as $\frac{-\beta}{\lambda_1}\int\limits_0^p q \partial_q^2 g(q) \mathrm{d}q$.
In particular, for the Boltzmann-Gibbs entropy $g_{\rm BG}(q)=-q\ln q$ and $\beta=\lambda_1$, yielding $\partial_q^2 g_{\rm BG}(q)=-q^{-1}$, which results in $F_{\rm BG}[p]=p$. Since $F[p]$ is formulated based on $g(p)$ in ($\ref{gen_F}$), we call the resulting equation 
\begin{equation}
\partial_t p=D\left\{\beta \partial_x(p\partial_x u )+\partial_x^2\left( \frac{-\beta}{\lambda_1}\int^p q \partial_q^2 g(q) \mathrm{d}q\right)\right\}
\label{gnewFPdef}
\end{equation} 
as the $g$-Fokker-Planck equation in order to distinguish it from the many other possible generalizations of FPEs. Note that in general many possible dynamics can lead to the same stationary state. However, as our derivation shows, if the dynamics, given by a non-linear Fokker-Planck equation, is constrained to be in the form of (\ref{gFPdef}), then $F[p]$ is determined by the stationary state, or, equivalently, by $g$, up to an additive constant.

In the following, let us consider the $g$-Fokker-Planck equation with no external potential, 
\begin{equation}
\partial_t p(x,t) = D \partial_x^2 F[p(x,t)].
\label{gFP_no_potential}
\end{equation}
We assume that the solution of (\ref{gFP_no_potential}) exists, at least from an appropriate initial condition, and it reaches the asymptotic limit $p(x,t)\rightarrow 0$ for all $x$.
In this asymptotic limit, the scaling rule (\ref{anom_diff_0}) applies to $p(x,t)$. Thus, if we change to the rescaled variables $x'=x/\tau^{\gamma}$ and $t'=t/\tau$, the derivatives according to the new variables can be written as 
\begin{equation}
\partial_t=\tau^{-1}\partial_{t'}, \;\;\; \partial_x^2=\tau^{-2\gamma}\partial_{x'}^2.
\label{derriv}
\end{equation}
Using (\ref{anom_diff_0}) and (\ref{derriv}), the $g$-Fokker-Planck equation with no external potential given in (\ref{gFP_no_potential}) in the rescaled variables can be formulated as
\begin{equation}
\partial_t p(x,t)=\tau^{-1}\partial_{t'}\left(\tau^{-\gamma}p(x',t')\right)=\tau^{-(\gamma+1)}\partial_{t'}p(x',t')=D\partial_x^2F[p(x,t)]=\tau^{-2\gamma}D\partial_{x'}^2\left(F[\tau^{-\gamma}p(x',t')]\right).
\label{general_FP_scaling}
\end{equation}
The $F[\tau^{-\gamma}p(x',t')]$ term on the right hand side can be further transformed based on the scaling of $g(p)$ given in (\ref{g_scaling}), where by a change of variable $\tilde{q}=q\tau^{\gamma}$ we obtain $g(\tilde{q}\tau^{-\gamma})=g(\tilde{q})\tau^{-\gamma c}$ (being valid for $\tilde{q} << 1$).
By substituting this into (\ref{gen_F}) we obtain
\begin{equation}
F[\tau^{-\gamma}p(x',t')]=\frac{-\beta}{\lambda_1}\int_0^{\tau^{-\gamma}p(x',t')}q \partial_q^2g(q)\mathrm{d}q=\frac{-\beta}{\lambda_1}\int_0^{p(x',t')}\tau^{-\gamma}\tilde{q}\tau^{2\gamma}\partial_{\tilde{q}}^2g(\tilde{q})\tau^{-\gamma c}\tau^{-\gamma}\mathrm{d}\tilde{q}=\tau^{-\gamma c}F[p(x',t')].
\label{general_F_scaling}
\end{equation}
According to that, the $g$-Fokker-Planck equation (\ref{general_FP_scaling}) in the rescaled variables yields
\begin{equation}
\tau^{-(\gamma+1)}\partial_{t'}p(x',t')=\tau^{-2\gamma-\gamma c}D\partial_{x'}^2 F[p(x',t')].
\end{equation}
Consequently, the dynamics remains to be governed by the original g-Fokker-Planck equation (\ref{gFP_no_potential}) for any $\tau$ if and only if the prefactors of both sides are equal for any $\tau$, that is, the exponents must coincide,
\begin{equation}
-(\gamma+1)=-2\gamma-\gamma c.
\label{eq_gamma}
\end{equation}
By rearranging (\ref{eq_gamma}) we obtain the main result of the paper
\begin{equation}
\gamma=\frac{1}{1+c},
\label{eq:main_result}
\end{equation}
providing a general relation 
between the exponent $\gamma$ related to the anomalous diffusion, characterizing the scaling of $p(x,t)$ in the  $p(x,t)\rightarrow 0$ limit and the exponent $c$, describing the scaling of the generalized entropy with the phase space volume. 
Note that the derivation above only requires $g(p)$ to obey the asymptotic scaling relation (\ref{g_scaling}) and the existence of the solution to (\ref{gFP_no_potential}) with the asymptotic limit $p(x,t)\rightarrow 0$ reached for all $x$.

\begin{figure}
  \centering
    \includegraphics[width=1.0\textwidth]{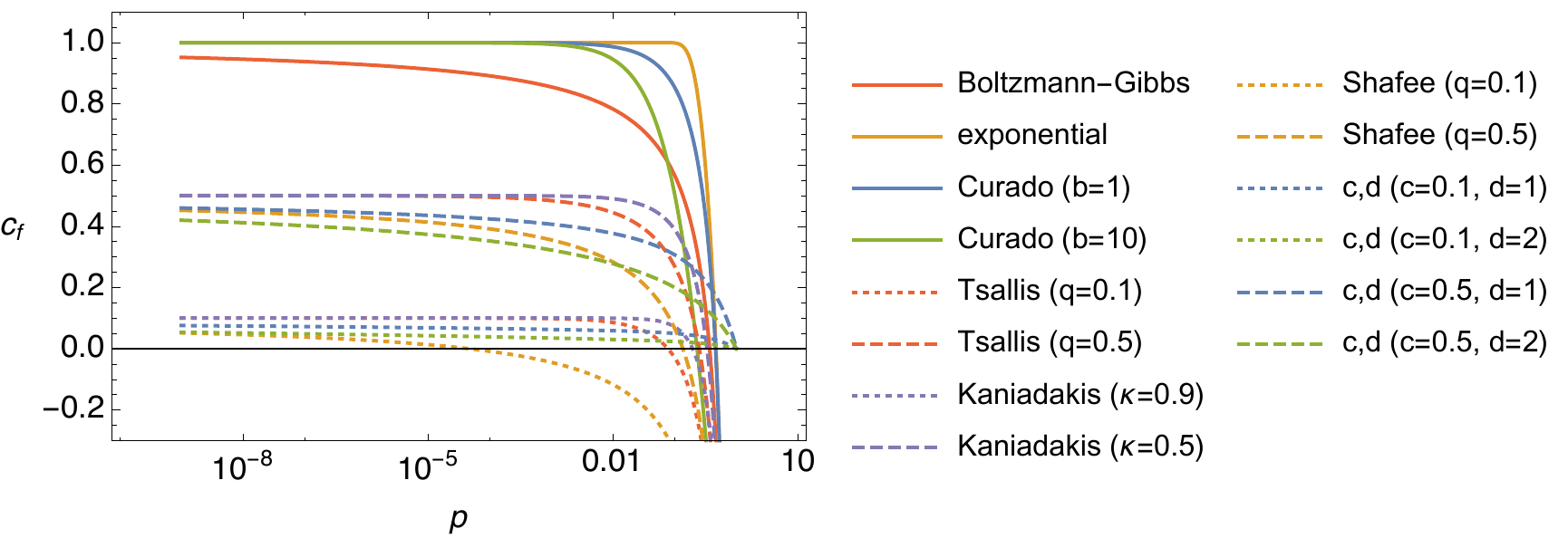}
      \caption{Finite scale phase space volume scaling exponent $c_f(p)$ of some generalized entropies belonging to asymptotic classes $c=0.1$, $c=0.5$ and $c=1$. According to the rigorous formulation of the asymptotic phase space volume scaling, given by (\ref{g_scaling}), $g(p)\sim p^c$ when $p\rightarrow 0$. Based on this, we define $c_f(p)$ as the slope of the tangent of $g(p)$ on a log-log plot, $c_f(p)=\frac{\partial \ln g(p)}{\partial \ln p}=p\frac{\partial \ln g(p)}{\partial p}$. The convergence of the curves in the low $p$ regime indicates that the effects of the phase space volume scaling (which is illustrated in the left panel of Fig.\ref{fig:summary}.) are apparent already at finite scales. } 
      \label{fig:c_finite}
\end{figure}

In order to demonstrate this general result, in Table \ref{tab:examples}. we list a few different generalized entropy forms from the literature together with the corresponding $g$-Fokker-Planck equations and the related $\gamma$ and $c$ exponents. Although the actual algebraic form of the entropies along with their phase space volume scaling, their MaxEnt distributions and the corresponding generalized Fokker-Planck equations are different for any $c$, their asymptotic anomalous diffusion scaling is completely determined by $c$ via (\ref{eq:main_result}). This exemplifies the fact that although the mapping between entropies and Fokker-Planck equations are defined at any (phase space volume or time) scale, any entropy, characterized by asymptotic exponent $c$, can only be mapped to a Fokker-Planck equation describing anomalous diffusion with asymptotic exponent given by (\ref{eq:main_result}). In close connection to Table \ref{tab:examples}., Fig.\ref{fig:c_finite}. shows the finite scale phase space volume scaling of some generalized entropies, illustrating the numerous possible ways of convergence to the asymptotic value $c$.

\begin{table}
\begin{center}
\begin{adjustbox}{width=1\textwidth}
  \begin{tabular}{|c|c|c|c|c|c|}
  \hline
  entropy & $g(p)$ & $\Lambda(p)$ & g-Fokker-Planck equation & $c$ & $\gamma$ \\
    \hline
    \hline
    Boltzmann-Gibbs & $-p \ln p$ & $\ln p$ & $\partial_t p=\partial_x^2 p$ & $1$ & $\frac{1}{2}$   \\ \hline
    exponential \cite{tsekouras2005generalized} & $p\left(1-e^{\frac{p-1}{p}}\right)$ & $e^{\frac{p-1}{p}}\left(\frac{p+1}{p}\right)-2$ & $\partial_t p=\partial_x^2 e^{\frac{p-1}{p}}$ & $1$ & $\frac{1}{2}$\\
    \hline
   Curado \cite{curado2004stability} & $1-e^{-bp}+p\left(e^{-b}-1 \right ),~ b>0$ & $\frac{1-e^{b(1-p)}}{b}$ & $\partial_t p=\partial_x^2 \left[1 -e^{-bp}(bp+1)\right]$ & $1$ & $\frac{1}{2}$ \\   \hline
    Tsallis \cite{tsallis1988possible} & $\frac{p-p^q}{q-1},~ 0<q\leq 1$ & $\frac{1-p^{q-1}}{1-q}$  & $\partial_t p=\partial_x^2 p^q$ & $q$ &  $\frac{1}{1+q}$  \\ \hline
    Kaniadakis \cite{kaniadakis2002statistical} & $-\frac{p^{1+\kappa}-p^{1-\kappa}}{2\kappa},~ 0<\kappa<1$ & $\frac{(1+\kappa)p^{\kappa}-(1-\kappa)p^{-\kappa}}{2\kappa}-1$ & $\partial_t p=\partial_x^2 [p^{1+\kappa}+p^{1-\kappa}]$ & $1-\kappa$ & $\frac{1}{2-\kappa}$ \\ \hline
    Shafee \cite{shafee2007lambert} & $-p^q \ln p,~ 0<q\leq 1$ & $\frac{1-p^{q-1}(1+q\ln p)}{1-2q}$ & $\partial_t p=\partial_x^2 [p^q(1+(q-1)\ln p)]$ & $q$ & $\frac{1}{1+q}$ \\
    \hline
 \multirow{2}{*}{$c,d$ \cite{hanel2011comprehensive}} & $er\Gamma\left(1+d,1-c\ln p\right)-cr,$ & \multirow{2}{*}{$r\left[1-\left(1-c\ln p\right)^d p^{c-1} \right ]$} &$\partial_t p=\partial_x^2\Bigl [ d\Gamma\left(d,1-c\ln p\right)$ & \multirow{2}{*}{$c$} &\multirow{2}{*}{ $\frac{1}{1+c}$} \\
  &  $r=\frac{1}{1-c+cd},~0<c\leq 1,~d\in\mathbb{R}$  & & $-\left(1-\frac{1}{c} \right )\Gamma\left(1+d,1-c\ln p\right) \Bigr ]$ & & \\   
    \hline
    \end{tabular} 
    \end{adjustbox}
  \caption{Generalized entropies $S=\int g\left(p(u)\right) \mathrm{d}u$, the inverse of their MaxEnt distribution $\Lambda(p)$, their corresponding g-Fokker-Planck equation (with no external potential), their phase space scaling exponent $c\in (0,1]$ and their anomalous diffusion scaling exponent $\gamma$. Different entropies with the same exponent $c$ might have different algebraic forms, different MaxEnt distributions and different corresponding Fokker-Planck equations, but the asymptotic anomalous diffusion scaling exponent $\gamma$ is always the same. Note that for the sake of comparison, $\lambda_0$ and $\lambda_1$ in $\Lambda(p)=\frac{g'(p)-\lambda_0}{\lambda_1}$ are set by the conditions $\Lambda(1)=0$ and $\Lambda'(1)=1$. Also the factor prior to $\partial_x^2$ are set to $1$ for simplicity which can be understood as the time being appropriately rescaled.}
  \label{tab:examples}
\end{center}
\end{table}

\section*{Discussion}


In this paper, we considered a class of stochastic processes, describing systems possibly composed of strongly interacting units or governed by non-Markovian dynamics, which can be macroscopically modelled by non-linear Fokker-Planck equations in the form of (\ref{gFPdef}).  These equations generalize the linear Fokker-Planck equation by replacing the probability density $p(x,t)$ in the diffusive term $\partial_x^2 p(x,t)$ by an effective density $F[p(x,t)]$. The actual form of $F[p]$ determines both the stationary and the non-stationary behaviour of the process. The non-stationary behaviour in the presence of no external potential, i.e., the solution of (\ref{gFP_no_potential}), can be classified according to the spread of the probability density by the anomalous diffusion scaling exponent $\gamma$, defined by eq. (\ref{anom_diff_0}). This exponent $\gamma$ provides a measure of deviation from ordinary diffusion, characterized by $\gamma=1/2$.

Another macroscopic approach of modelling stochastic processes is through the construction of generalized entropy functionals $S$ which are maximized by the stationary state of the processes. As it has been already shown, a meaningful classification of generalized entropies over a discrete phase space indexed by $i$, $S[p]=\sum_{i=1}^W g(p_i)$, can be given by their scaling with phase space volume $W$, characterized asymptotically by the exponent $c$. 
Here we consider analogous continuous entropy functionals in the form of $S[p(u)]=\int g\left(p(u)\right ) \mathrm{d}u$. Similarly to the exponent $\gamma$ regarding the non-stationary regime, the phase space volume scaling exponent $c$ of the entropy quantifies the deviation from uncorrelated, Markovian systems (characterized by $c=1$) at the stationary regime.

The two approaches, one based on Fokker-Planck equations and the other on entropies, are consistent at the stationary regime if the stationary solution of the Fokker-Planck equation equals to the maximum entropy distribution according to the generalized entropy.
In this paper we show that this consistency criterion implies that asymptotically, i.e., at $p\to 0$, the anomalous diffusion scaling exponent $\gamma$ and the phase space volume scaling exponent of the entropy $c$ bijectively determine each other via the relation $\gamma=\frac{1}{1+c}$. 
Asymptotically, this result generalizes that of Tsallis and Bukman \cite{tsallis1996anomalous}, now being valid for any generalized entropy functional satisfying some general asymptotic conditions. In addition, since the explicit solution of the corresponding Fokker-Planck equation might either not be available, or possibly have infinite variance, our derivation do not rely on the computation of any of these.
Our results suggests that either of the asymptotic exponents $\gamma$ and $c$ is indeed providing a useful characterization of the systems themselves, and not just describing their behaviour in the stationary or non-stationary regime.
Furthermore, the surprising versatility of the theoretical framework behind Tsallis statistics to model various aspects of strongly interacting systems might be explained by the fact that the family of Tsallis entropies, characterized by their deformation index $q$, provides an algebraically simple representative of each such asymptotic equivalence class.


\end{document}